\documentclass[a4paper, 10pt, conference]{ieeeconf}
\usepackage{graphicx}
\usepackage{algorithm2e}

\IEEEoverridecommandlockouts
\title{\LARGE \bf
    Towards an Angry-Birds-Like Game System for Promoting Mental Well-Being of Players Using Art-Therapy-Embedded\\ Procedural Content Generation
}
\author{
    Zhou Fang$^{1}$, Pujana Paliyawan$^{2}$, Ruck Thawonmas$^{1}$ and Tomohiro Harada$^{1}$\\
    $^{1}$College of Information Science and Engineering\\
    $^{2}$Research Organization of Science and Technology\\
    Ritsumeikan University, Japan\\
    ruck@is.ritsumei.ac.jp
}
\begin{document}
\maketitle
\thispagestyle{empty}
\pagestyle{empty}

\begin{abstract}
This paper presents an integration of a game system and the art therapy concept for promoting the mental well-being of video game players. In the proposed game system, the player plays an Angry-Birds-like game in which levels in the game are generated based on images they draw. Upon finishing a game level, the player also receives positive feedback (praising words) toward their drawing and the generated level from an Art Therapy AI. The proposed system is composed of three major parts: (1) a drawing recognizer that identifies what object is drawn by the player (Sketcher), (2) a level generator that converts the drawing image into a pixel image, then a set of blocks representing a game level (PCG AI), and (3) the Art Therapy AI that encourages the player and improves their emotion. This paper describes an overview of the system and explains how its major components function.

\end{abstract}


\section{INTRODUCTION}

Art therapy \cite{handbook-of-a-t}, also known as art psychotherapy, is a method to reduce pressure and improve mental well-being. Not as long as other therapies, the term ``art therapy" was coined by Adrian Hill in 1942. Although art therapy is a relatively young therapeutic discipline, it shows a potent effect on promoting mental well-being without any specific medical instruments. Commonly, art therapy is used for patients who need to release uneasy emotions. However, according to the literature, art therapy also worked with non-patients and helped them reduce their mental pressure.

\begin{figure}[htbp]
\centering
\includegraphics[width=0.4\textwidth]{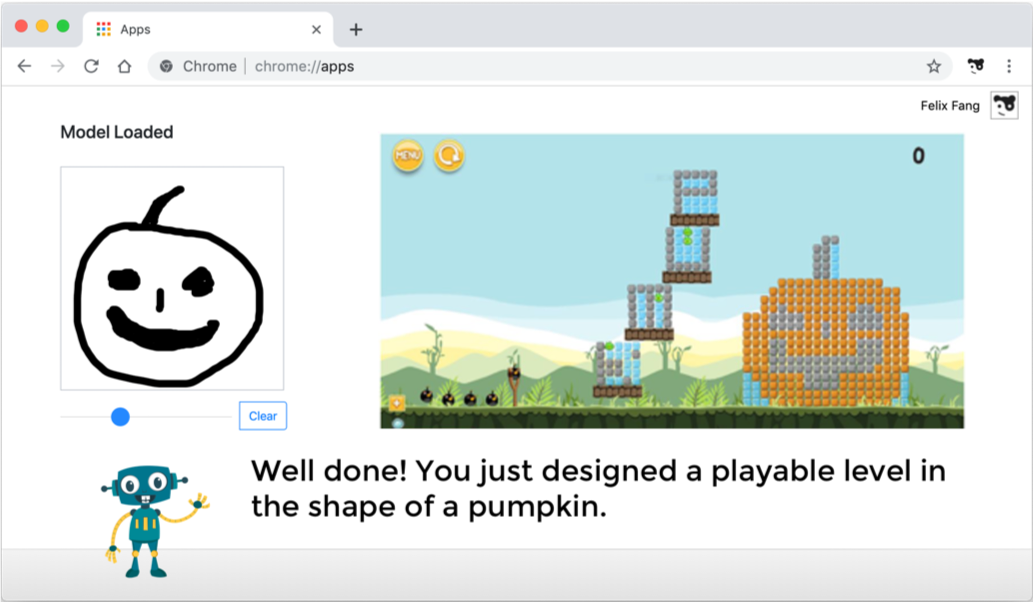}
\caption{Interface of the Proposed Game System}
\label{system-demo}
\end{figure}

Many games nowadays could bring benefits of various kinds to players, including health-related benefits~\cite{d1}. One example of games for promoting mental well-being was presented by Xu et al.~\cite{c4} where an Angry-Bird-like game was used to encourage smiles and consequently the positive affect of players. With the belief that video games can be combined with the art therapy concept to enhance the mental well-being of people, this paper presents a design of an Angry-Bird-like game system that employs the concept of art therapy for promoting mental well-being (Fig.~\ref{system-demo}).

\begin{figure}[htbp]
\centering
\includegraphics[width=0.38\textwidth]{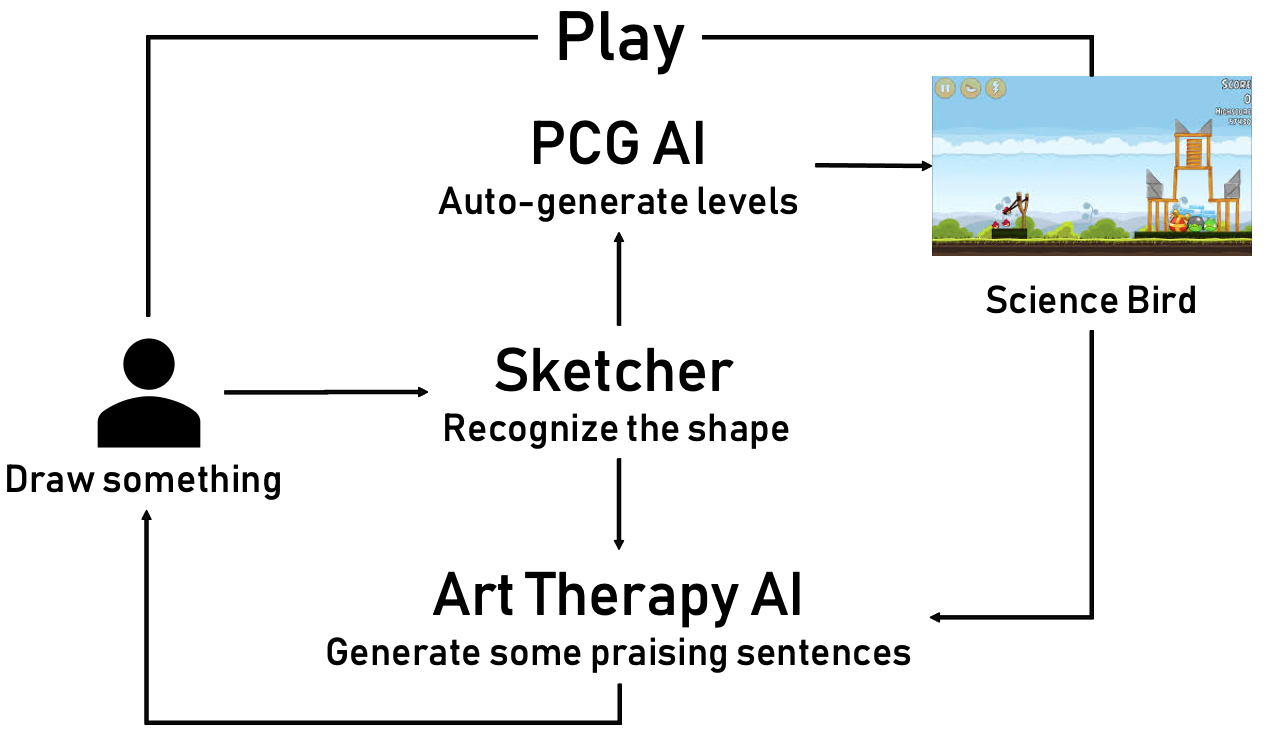}
\caption{Overview of the Proposed Game System}
\label{system-overview}
\end{figure}

\section{EXISTING WORK}

\subsection{Procedural content generation (PCG)}

Procedural content generation (PCG)~\cite{PCG} in games is to create game contents, such as game stages or levels, procedurally using algorithms. It is relatively common and has been used in a lot of popular games during recent years. A well-known PCG competition is the level generation competition for Angry Bird annually held by IEEE Conference on Games (CoG) since 2016 (then IEEE Conference on Computational Intelligence and Games).

\subsection{Angry Bird for Health Promotion}

Angry Bird has been a popular platform for game research. There is a clone game of it called Science Birds~\cite{science-birds-link}, which is open-source and widely used in academia. Level generation is an example of PCG. Here, PCG is used to generate levels.

Most of the previous studies on Angry-Bird-like games level generation were not directly focused on promoting players' emotion (\cite{c4}). On the contrary, Yang et al.~\cite{c3} presented and investigated two methods designed for encouraging players to smile during gameplay. A follow-up study by Xu et al.~\cite{c4} proposed a game system that requires the player to smile in order to erase fog covering the game level. In addition, Xu et al. also proposed ``Pixel Image Level Generator(PILG)" for generating game levels based on the input pixel image. Results in these previous studies showed that the promotion of mental well-being using an Angry-Bird-like game is promising.

In contrary to the previous studies above, instead of smiling, we use drawing to receive input from the player. In addition, Yang et al.'s level generator~\cite{c3} only focused on placing TNT blocks or adding a fog to randomly generated game levels, while Xu et al.'s PILG~\cite{c4} could generate the whole level, but the generator was for building some difficult levels---in the latter work, it was noted that the player cannot clear certain levels. Differently, we aim at generating a whole level and take into consideration that the player may have negative reactions if they could not clear the levels. We want to ensure that generated levels are not too difficult.

\subsection{Sketcher}

Sketcher~\cite{sketcher-link} is an open-source tool that uses Convolutional Neural Networks (CNN) to recognize drawings. Sketcher is run on TensorFlow.js, and it works with any modern web browser. With this tool, when the player draws something in the canvas inside the web page, Sketcher would recognize it and return names of five most similar objects with confidence rates. This study applies Sketcher for receiving drawing data as input for level generation.

\section{PROPOSED GAME SYSTEM}
An overview of our system is shown in Fig.~\ref{system-overview}. There are three major modules: PCG AI, Sketcher, and Art Therapy AI. Sketcher is explained in the previous section. In this section, the other two are introduced in the subsections below, respectively.

\subsection{PCG AI}

\begin{figure}[b]
\centering
\includegraphics[width=0.45\textwidth]{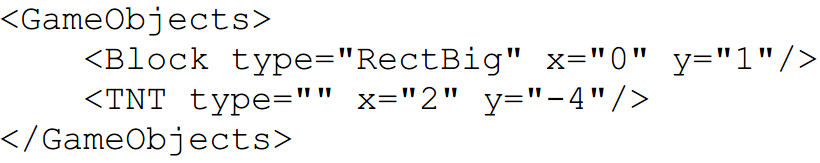}
\caption{Level File Example}
\label{level-xml}
\end{figure}

The level of Science Bird is written as an XML file, as shown in Fig.~\ref{level-xml}.  Algorithm~\ref{level-generator-algo} works as follows. First, it reads a pixel image drawn by the player, and open a blank level file, to which a generated level will be saved. Second, it binarizes and converts the image to an array containing pixel data. Then, it adds block information in the level file based on the pixel data.

Note that in order to build a level with a high stability (i.e., the level should not collapse or shake before the player shoots it), we need to fill blocks in empty spaces. Moreover, in order to create a level that is not too difficult, one of the strategies is to  randomly convert some blocks in the generated level into TNT blocks.

\begin{algorithm}[h]
\SetAlgoLined
GetImageAndOpenLevelFile()\;
ImageBinarization()\;
\While{ColumnNumber $\leq$ MaxColumnNumber}{
    LastBlock $\gets$ 0\;
    \While{PixelNumber $\leq$ MaxPixelNumber}{
        \If{IsBlack()}{
            \If{PixelNumber - LastBlock $\geq$ 2}{
                FillBlocksOut()\;
            }
            AddBlock()\;
            LastBlock $\gets$ PixelNumber\;
        }
    }
}
ConvertBlockToTNTRandomly()\;
\label{level-generator-algo}
\caption{Level Generator}
\end{algorithm}

\subsection{Art Therapy AI}

Art Therapy AI aims at generating praising phases, as doctors do in traditional art therapy. It uses the result image recognition in Sketcher and the data from gameplay to make the feedback more human-like. For example, if the player cannot clear the level, it would send ``Good job! You just designed a hard level in the shape of a smiling face."

\section{CONCLUSIONS}

Design of an Angry-Bird-like game system embedded an art therapy concept for promoting the mental well-being of players is presented. The system allows the player to play game levels generated by the image they draw, and also get positive feedback towards their drawing. To the best of our knowledge, this is the first combination of art therapy and a video game. Our future work includes full-scale experiments to evaluate the effectiveness of this system.

\section*{Acknowledgement}
This research was supported in part by Grant-in-Aid for Scientific Research (C), Number 19K12291, Japan Society for the Promotion of Science, Japan.

\addtolength{\textheight}{-12cm}

\end{document}